\documentclass[twocolumn,showpacs]{revtex4}%
\usepackage{epsfig}
\usepackage{amsmath}
\usepackage{graphicx}
\usepackage{amsfonts}
\usepackage{amssymb}%
\begin{document}
\preprint{TPJU}
\title{On the exotic eikosiheptaplet}
\author{Michal Praszalowicz}
\email{michal@if.uj.edu.pl}
\affiliation{M. Smoluchowski Institute of Physics, Jagellonian University, ul. Reymonta 4,
30-059 Krak{\'o}w, Poland}
\author{Klaus Goeke}
\email{Klaus.Goeke@tp2.ruhr-uni-bochum.de}
\affiliation{Institut f\"ur Theoretische Physik II, Ruhr-Universit\" at Bochum, D--44780
Bochum, Germany}
\date{\today }

\begin{abstract}
We use the chiral quark soliton model to estimate masses and widths of the two
eikosiheptaplets (27-plets of SU(3) flavor) of spin 3/2 and 1/2 that emerge in
the rigid rotator quantization. We use as input: hyperon decays, $\Theta^{+}$
mass and width. While $27_{3/2}$ has small widths (although much larger than
the values allowed by the partial wave analysis), $27_{1/2}$ has large decay
widths to antidecuplet. However exactly for this decay channels the widths are
suppressed in the large $N_{c}$ limit.

\end{abstract}

\maketitle








\section{Introduction}

One of the most puzzling results of the chiral quark-soliton model ($\chi$QSM)
for exotic baryons consists in a very small hadronic decay width, governed by
the decay constant $G_{\overline{10}}$. While the small mass of exotic states
is rather generic for all chiral models \cite{BieDot,prasz,DPP} the smallness
of the decay width appears as a subtle cancelation of three different terms
\cite{DPP} that contribute to $G_{\overline{10}}$. We are therefore trapped
between two extremes. On one hand $\Delta$ decay width which is suppressed in
large $N_{c}$ limit  is numerically rather large, above 100 MeV, on the other
hand $\Theta^{+}$ decay width which scales like $N_{c}^{0}$, is numerically
tiny, below 1 MeV. If narrow pentaquarks exist, large $N_{c}$ argument is not
enough to claim consistency of the model, some degree of cancelation in the
decay coupling is needed. In this paper we investigate this problem for the
next exotic SU(3) representation, namely 27-plet, called eikosiheptaplet.

Following the prescription of Adkins, Nappi and Witten \cite{ANW} (criticized
recently in Ref.\cite{Weigel:2007yx}) the decay width in solitonic models is
calculated in terms of a matrix element $\mathcal{M}$ of the collective axial
current operator corresponding to the emission of a pseudoscalar meson
$\varphi$:
\begin{align}
\hat{O}_{\varphi}^{(8)} &  =3\times\text{const.}\times p_{\varphi}%
^{i}\label{operator}\\
&  \times\sum_{i=1}^{3}\left(  a_{1}\,D_{\varphi i}^{(8)}+a_{2}\,d_{ibc}%
\,D_{\varphi b}^{(8)}\hat{S}_{c}+\frac{a_{3}}{\sqrt{3}}\,D_{\varphi8}%
^{(8)}\hat{S}_{i}\right)  \nonumber\\
&  =3\times p_{\varphi}^{i}\nonumber\\
&  \times\sum_{i=1}^{3}\left(  G_{0}\,D_{\varphi i}^{(8)}-G_{1}\,d_{ibc}%
\,D_{\varphi b}^{(8)}\hat{S}_{c}-\frac{G_{2}}{\sqrt{3}}\,D_{\varphi8}%
^{(8)}\hat{S}_{i}\right)  \nonumber
\end{align}
where in the last line of eq.(\ref{operator}) we have displayed the operator
in the form often used in the literature. Here $D_{\varphi i}^{(8)}$ are SU(3)
Wigner matrices, $\hat{S}_{i}$ collective spin operator, $p_{\varphi}^{i}$
meson momentum (for more details on the collective quantization and baryon
wave functions see \emph{e.g.} Ref.\cite{Blotzetal}). Constants $a_{1,2,3}$
are constructed from the so called \emph{moments of inertia} that are
calculable in $\chi$QSM \cite{spinp, Wakamatsu}. A multiplicative constant has
to be fixed from the generalized Goldberger-Treiman relation \cite{DPP,YKG}.
Alternatively, following the \emph{model independent approach} of Adkins and
Nappi \cite{AN}, one can treat $a_{1,2,3}$ as \emph{free} constants and try to
extract their phenomenological values from the hyperon decays \cite{spinp,YKG}.

The predictive power of the model independent approach for exotic baryons is,
however, hampered by the fact that only one linear combination constructed
from two free parameters $a_{1,2}$, namely
\[
a_{1}-\frac{1}{2}a_{2},
\]
enters the hyperon decay widths, whereas for the decay widths of exotic states
both $a_{1}$ and $a_{2}$ are needed separately. The same problem occurs for
baryonic masses \cite{prasz,DPP} where no information on the exotica can be
retrieved from the regular baryon spectra alone (and similarly for magnetic
moments \cite{newmag}).

One is therefore forced to introduce some additional assumptions to fix the
remaining coefficient. In the original work of Ref.\cite{DPP} masses were
fixed by a requirement that nucleon resonance N$^{\ast}(1710)$ was a member of
antidecuplet. Decay widths were estimated with the help of hyperon
semileptonic decays and $g_{\pi\text{NN}}$ used as an input with some other
simplifying assumptions. A complete phenomenological analysis in this spirit
can be found in Ref.\cite{EKP}.

Another possibility to constrain the undetermined parameter is to go beyond
the SU(3) symmetry limit and include higher order symmetry breaking terms
\cite{YKG}. Why going off the symmetry limit may be at all of help? The answer
is very simple: the baryonic wave functions belong no longer to pure SU(3)
multiplets, but contain $m_{s}$ dependent admixtures of higher
representations. For example a nucleon contains an admixture of antidecuplet
cryptoexotic nucleonic state. As a result, the matrix element of any operator
(\emph{e.g.} the decay operator (\ref{operator})) contains -- apart from the
leading term -- exotic transitions from antidecuplet to octet as a nonleading
correction. By fitting the decay rates with $m_{s}$ corrections one is
therefore able to constrain the otherwise undetermined parameter.

The first estimate of the $\Theta^{+}$ mass in the Skyrme model has been done
in this way already long ago \cite{prasz}. More recently magnetic moments
\cite{newmag} and $\Theta^{+}$ decay width \cite{YKG} have been evaluated by
applying the above mentioned procedure. There all higher representations are
treated as stable hadronic states, rather than as wide resonances. In
particular admixture of eikosiheptaplet (27-plet) is here of importance (see
Fig.1 of Ref.\cite{MPbreaking}). Indeed the contamination of baryonic wave
functions by eikosiheptaplet reaches 20-30\% .

It is therefore of importance to check whether the eikosiheptaplet may be
indeed considered as a (semi) stable exotic representation. Not only can it
mix with ordinary baryons, but it contains a number of exotic states that may
be of interest by themselves, the isotriplet of $\Theta$ states being the most
prominent example. Since transitions to exotic representations enter through
representation mixing which itself is of the order of $m_{s}$, (semi)
stability of eikosiheptaplet has to be valid in the leading order of
perturbartive expansion in the strange quark mass. Therefore in our analysis
of the decay widths we work in the chiral limit.

In chiral models all baryon representations have positive parity and spin
corresponding to the isospin of states with $Y=1$. For eikosiheptaplet that
means that we have two distinct representations, one of spin 3/2 and the
second one of spin 1/2 (\emph{i.e.} $27_{3/2}$ and $27_{1/2}$ respectively),
the latter being heavier. In this work we shall concentrate on the lightest
states, namely on the isospin triplet of $\Theta_{27}$, on $\Delta_{27}$ of
isospin 3/2 and on N$_{27}$ states of isospin 1/2. These states are light (for
$27_{3/2}$) and have been looked for in various experiments. Apart from still
unconfirmed reports by STAR \cite{star_thetapp} recent partial wave (PW)
analysis of meson-nucleon scattering data put stringent limits on possible
existence of $\Theta_{27}$ and $\Delta_{27}$ states \cite{Azimov}. These
states may be incorporated into the PW analysis provided that their widths are
of the order of tens keV. As we shall see $\chi$QSM predicts that their widths
are order of magnitude larger. Although still small on a hadronic scale, they
are much too large to be accommodated by PW analysis.

Throughout this paper we shall assume that $\Theta^{+}$ exists with mass 1535
MeV and width smaller than 1 MeV. This input allows us to constrain all models
parameters except $\Sigma_{\pi\text{N}}$. If additionally we assume that
$\Xi_{3/2}$ has mass $\sim$1860 MeV, also pion-nucleon sigma term is fixed
$\mathit{\Sigma}_{\pi\text{N}}=73$ MeV.

There have been already a few calculations in the literature of the
eikosiheptaplet masses and widths in chiral soliton models \cite{Borisyuk1}%
\nocite{Ma1,Borisyuk2,Ma2,Weigelmon,Josip,Ma3}--\cite{Kopel}. In this paper we
use the mass formula of Ref.\cite{Borisyuk2}. Generically the mass of the
lowest $I=1$ multiplet of $\Theta_{27}$ states in $27_{3/2}$ is almost
degenerate with $\Theta^{+}$ of $\overline{10}$. On the contrary, $27_{1/2}$
is substantially heavier.

As far as widths are concerned our calculations differ in three aspects from
those of Refs.\cite{Borisyuk1}\nocite{Ma1,Borisyuk2}--\cite{Ma2}. Firstly, in
Ref.\cite{Borisyuk1} one considers only the leading $G_{0}$-term, whereas in
Refs.\cite{Ma1,Borisyuk2,Ma2} the constant $G_{2}$ has been neglected. Indeed,
$G_{2}$ (or more precisely $a_{3}$) is small as it is directly related to the
singlet axial current. Even though it is really small, it can be safely
neglected \emph{only} if there is no cancelation between $G_{0}$ and $G_{1}$,
so that the pertinent linear combination of $G_{0}$ and $G_{1}$ is much larger
than $G_{2}$ itself. In the decays of antidecuplet, $27_{3/2}\rightarrow
8+$meson and $27_{1/2}\rightarrow10+$meson strong cancelations are indeed
present and $G_{2}$ cannot be neglected. In this paper we use $a_{3}$
extracted from the chiral limit fits to the semileptonic hyperon decays that
is definitely not consistent with zero. Secondly, we use the
Goldberger-Treiman relation to fix the constant entering eq.(\ref{operator}),
so that $G_{0,1,2}$ depend on the decay in question, whereas in
Refs.\cite{Borisyuk1}\nocite{Ma1,Borisyuk2}--\cite{Ma2} they were considered
as universal. Thirdly, instead of calculating the decay widths and masses for
a fixed choice of model parameters we explore the residual freedom within the
model and calculate the \emph{range} of values, rather than only one number.
Finally some calculations \cite{Ma2} took partially into account the effects
of the symmetry breaking which is neglected in our paper.

We show that $27_{3/2}$ is in a sense "well behaving" having small widths to
octet with most other channels kinematically suppressed. On the contrary,
$27_{1/2}$ has large decay widths to antidecuplet, with small decay widths to
other channels. However, precisely in the case of $27\rightarrow\overline{10}
$ transition the phase space is formally suppressed in the large $N_{c}$
limit. The situation reminds the decay of $\Delta$ and $\Theta^{+}$, the first
one being numerically large, but formally damped in the large $N_{c}$ limit
with the second one being numerically small but $\mathcal{O}(1)$ as far as
$N_{c}$ counting is concerned.

The paper is organized as follows. In Section 2 we give an overview of the
nonrelativistic formalism to calculate the decay widths using the generalized
Goldberger-Treiman relation. We fix two out of three axial constants and
define model parameters. Finally we calculate the masses of antidecuplet and
eikosiheptaplet. In Section 3 we express antidecuplet and decuplet amplitudes
entering the decay widths through couplings $G_{10}$ and $G_{\overline{10}}$
and the SU(3) isoscalar factors. By fixing $\Theta^{+}$ decay width to be
below 1 MeV we constrain the axial coupling parameter space and give results
for the decay widths of other members of antidecuplet. In Section 4 we repeat
the calculations from the preceding section for eikosiheptaplets of spin $3/2$
and $1/2$. We perform phenomenological analysis of the pertinent decay
couplings -- the analogs of $G_{10}$ and $G_{\overline{10}}$ -- and calculate
the decay widths. We summarize our findings in Sect. 5. Some useful
group-theoretical formulae are collected in the Appendix.

\section{General formalism}

Throughout this paper we shall use the nonrelativistic formula for the decay
width \cite{DPP,Weigelwidth}
\begin{equation}
\mathit{\Gamma}_{B\rightarrow B^{\prime}+\varphi}=\frac{1}{8\pi}%
\frac{p_{\varphi}}{M\,M^{\prime}}\overline{\mathcal{M}^{2}}=\frac{1}{8\pi
}\frac{p_{\varphi}^{3}}{M\,M^{\prime}}\overline{\mathcal{A}^{2}}.
\label{gammadef}%
\end{equation}
The \textquotedblleft bar\textquotedblright\ over the amplitude squared
denotes averaging over initial and summing over final spin and over isospin.
Anticipating linear momentum dependence of the decay amplitude $\mathcal{M}$
\begin{equation}
\mathcal{M}_{B\rightarrow B^{\prime}+\varphi}=\left\langle \mathcal{R}%
_{S^{\,\prime}}^{\prime},B^{\prime}\right\vert \hat{O}_{\varphi}%
^{(8)}\left\vert \mathcal{R}_{S},B\right\rangle
\end{equation}
we have introduced reduced amplitude $\mathcal{A}$ where the momentum of the
outgoing meson%
\begin{equation}
p_{\varphi}=\frac{\sqrt{(M^{2}-(M^{\prime}+m_{\varphi})^{2})(M^{2}-(M^{\prime
}-m_{\varphi})^{2})}}{2M} \label{mom}%
\end{equation}
has been factored out. Here $\mathcal{R}$ stands for the SU(3) representation
and $S$ for spin. In Ref.\cite{DPP} following the approach of
Ref.\cite{Samios} $MM^{\prime}$ in eq.(\ref{gammadef}) was replaced by
$(M+M^{\prime})^{2}/4$ and furthermore the additional factor $M/M^{\prime}$
was inserted to sum up certain kinematical effects. We will not make such
alterations in the following. Instead, we will apply the generalized
Goldberger-Treiman relation that allows to relate the axial constants
$a_{1,2,3}$ to the constants $G_{0,1,2}$ by means of the following relation
\cite{YKG}:%
\begin{equation}
G_{0}=-\frac{M+M^{\prime}}{3f_{\varphi}}a_{1},\;G_{1,2}=\frac{M+M^{\prime}%
}{3f_{\varphi}}a_{2,3} \label{GT}%
\end{equation}
where $M$ and $M^{\prime}$ stand for the baryonic masses involved in the decay
$B\rightarrow B^{\prime}+\varphi$ and $f_{\varphi}$ denotes pseudoscalar meson
decay constant in the normalization where $f_{\pi}=93$ MeV, $f_{K}=115$ MeV
and $f_{\eta}=1.2f_{\pi}$ \cite{PDG} (we neglect $\eta-\eta^{\prime}$ mixing).
The use of eq.(\ref{GT}) makes constants $G_{0,1,2}$ \emph{decay dependent} in
contrast to previous analysis where they were considered to be universal, with
possible modification of the formula for the width (\ref{gammadef}).

In contrast to the early exploratory works we now \emph{know for sure} that if
$\Theta^{+}$ exists it is light and its width is small. Therefore we use these
two pieces of information to constrain the mass and the decay width of
$\Theta^{+}$ for which we take $M_{\Theta}=1535$ MeV and $\mathit{\Gamma
}_{\Theta\rightarrow\text{N}+\text{K}}\sim1$ MeV. With these parameters fixed
we calculate the decay widths of decuplet, antidecuplet and eikosiheptaplet
and discuss uncertainties of our results coming from the $m_{s}$ corrections.
In this respect we differ from Ref.\cite{YKG} where $m_{s}$ corrections were
used -- as explained in the Introduction -- to constrain input parameters.

In order to fix the input parameters $a_{1,2,3}$ we use a fit from
Refs.\cite{spinp,spin8} where one uses two linear combinations of known
hyperon decays, that in $\chi$QSM are free of the linear $m_{s}$ corrections%
\begin{equation}
a_{1}-\frac{1}{2}a_{2}=-2.675,\;a_{3}=0.678. \label{inputa}%
\end{equation}
With these parameters one obtains:%
\begin{equation}
g_{A}^{(3)}=1.27,\;g_{A}^{(8)}=0.43,\;g_{A}^{(0)}=0.68.
\end{equation}
These values overshoot present experimental results, especially for
$g_{A}^{(0)}$ (that ranges between $0.15-0.35$ \cite{Bass}). It should be,
however, remembered that $g_{A}^{(0)}$ is sensitive to the corrections of
higher order in $m_{s}$ that pull it down with respect to the chiral limit
estimate (see Fig.2 in \cite{spin8}).

Let us stress that parametrization (\ref{inputa}) is theoretically very
appealing, because one does not need to refit leading order parameters
$a_{1,2,3}$ when $m_{s}$ corrections are included. Nevertheless the overall
quality of the fit is of course better when full formula with $m_{s}$
corrections is used \cite{YKG}. To check sensitivity of our results to the
fitting procedure, we have also used different set of parameters (that will be
called fit 2 in the following) which better fits $g_{A}^{(0)}$ in the leading
order:%
\begin{equation}
a_{1}-\frac{1}{2}a_{2}=-5.4,\;a_{3}=0.3 \label{inputa1}%
\end{equation}
which gives:%
\begin{equation}
g_{A}^{(3)}=1.27,\;g_{A}^{(8)}=0.36,\;g_{A}^{(0)}=0.3.
\end{equation}

Contenting ourselves with input parameters (\ref{inputa},\ref{inputa1}) we can
check our formalism against the hadronic data. Firstly, let us compute the
pion-nucleon coupling constant $g_{\pi\text{NN}}$ that for both fits reads%
\begin{equation}
g_{\pi\text{NN}}=\frac{7}{10}(G_{0}+\frac{1}{2}G_{1}+\frac{1}{14}G_{2})=12.8
\end{equation}
vs. experimental value of $13.1-13.3$ \cite{EKP}. Here the numerical result
has been obtained by putting $M=M^{\prime}=M_{\text{N}}$ in eq.(\ref{GT}).
Secondly, anticipating results of the next section, we can also quote our
prediction for the decay width of $\Delta$ obtained by means of
eq.(\ref{gammadef})%
\begin{equation}
\mathit{\Gamma}_{\Delta}=104\;(106)\,\text{MeV} \label{GDel}%
\end{equation}
in fair agreement with experiment (the number in parenthesis refers to the
parameters of eq.(\ref{inputa1})). Note that one may improve this result by
including a phenomenological factor $M_{\Delta}/M_{\text{N}}$
\cite{DPP,Samios} that would scale (\ref{GDel}) up to 134 MeV. Also $m_{s}$
corrections increase the $\Delta$ width (in this case by $25\%-30\%$
\cite{MPbreaking}).

For the decays of exotic states we have to know $a_{1}$ and $a_{2}$
separately. We therefore parameterize
\begin{equation}
a_{1}=\rho,\;a_{2}=5.352+2\rho,\,a_{3}=0.68. \label{inputa12}%
\end{equation}
It follows from the phenomenological analysis of Ref.\cite{YKG} that the
realistic range for $\rho$ lies within $-3$ to $-1.9$. In what follows we
shall fix $\rho$ to \emph{fit} the "experimental" width for $\Theta^{+}$. As
will be shown in eq.(\ref{Gamrho}), if we require $\mathit{\Gamma}_{\Theta}<1$
MeV then $\rho_{1}=-1.98<\rho<$ $\rho_{2}=-1.814$. For comparison we will also
use fit 2%
\begin{equation}
a_{1}=\rho,\;a_{2}=5.4+2\rho,\,a_{3}=0.3 \label{fit2}%
\end{equation}
varying $\rho$ within the limits $\rho_{1}=-1.933<\rho<\rho_{2}=-1.767$. All
numerical results in the following will be presented for fit 1 (\ref{inputa12}%
), modifications due the second choice of input parameters (\ref{fit2}) will
be discussed in Sect.V.

Finally, in order to use formula (\ref{gammadef}) we have to specify masses of
exotic states. To this end we parameterize all exotic masses in terms of one
parameter: $\mathit{\Sigma}_{\pi\text{N}}$, \emph{i.e.} the pion nucleon sigma
term that we will vary within the range of 40 -- 70 MeV.

In chiral quark soliton model baryon masses can be read off from the
collective hamiltonian%
\begin{align}
\hat{H}  &  =M_{cl}+\frac{1}{2I_{1}}S(S+1)\nonumber\\
&  +\frac{1}{2I_{2}}(C_{2}(\text{SU(3)})-S(S+1)-\frac{N_{c}^{2}}{12}) +
\hat{H}^{\prime} \label{Hamiltonian}%
\end{align}
where the symmetry breaking hamiltonian takes the following form:%
\begin{equation}
\hat{H}^{\prime}=\alpha D_{88}^{(8)}+\beta Y+\frac{\gamma}{\sqrt{3}}%
D_{8i}^{(8)}\hat{S}_{i}. \label{Hprim}%
\end{equation}
Matrix elements of $\hat{H}^{\prime}$ can be found $e.g.$ in
Refs.\cite{EKP,Borisyuk2}. For $M_{\Theta}=1535$ MeV the model parameters take
the following values (in MeV) as functions of $\mathit{\Sigma}_{\pi\text{N}}$
\cite{EKP,MPbreaking}:
\begin{equation}
\frac{1}{I_{2}}=152.4\,,\quad\frac{1}{I_{2}}=608.7-2.9\,\mathit{\Sigma}%
_{\pi\text{N}}\,. \label{I1I2}%
\end{equation}
and%
\begin{align}
\alpha &  =336.4-12.9\,\mathit{\Sigma}_{\pi\text{N}},\nonumber\\
\beta &  =-336.4+4.3\,\mathit{\Sigma}_{\pi\text{N}},\nonumber\\
\gamma &  =-475.94+8.6\,\mathit{\Sigma}_{\pi\text{N}} . \label{albega}%
\end{align}
Numerical results for antidecuplet obtained with the help of Eqs.(\ref{I1I2}%
,\ref{albega}) are summarized in Table \ref{a10mass}.

\begin{table}[h]
\caption{Masses of antidecuplet for different values of $\mathit{\Sigma}%
_{\pi\text{N}}$}%
\label{a10mass}
\begin{tabular}
[c]{|c|c|c|c|}\hline
$\mathit{\Sigma}_{\pi\text{N}}$ & 42~MeV & 55~MeV & 73~MeV\\\hline
$\Theta$ & 1535 & 1535 & 1535\\
N & 1709 & 1681 & 1642\\
$\Sigma$ & 1883 & 1827 & 1750\\
$\Xi_{3/2}$ & 2057 & 1974 & 1857\\\hline
\end{tabular}
\end{table}

Our choice for the values of $\mathit{\Sigma}_{\pi\text{N}}$ in Table
\ref{a10mass} is not accidental. For $\mathit{\Sigma}_{\pi\text{N}}=42$ MeV
the mass of the cryptoexotic nucleon resonance corresponds to the original
choice of \cite{DPP} who associated it with the known resonance N$^{\ast
}(1710)$. Almost for sure this choice is now ruled out, and this implies that
the new, narrow (as we will see below) nucleon resonance needs to be yet
discovered. There are several candidates for such states found both in partial
wave analysis \cite{Strakovsky}, $\eta$ photoproduction on nucleon (see
Ref.\cite{Kuznetsov} and references therein), and at STAR \cite{KabanaN}.
Next, the value of 55 MeV corresponds to $\mathit{\Sigma}_{\pi\text{N}}$
calculated within the model \cite{massigma}, and moreover it is the value for
which one of the symmetry breaking parameters (\ref{Hprim}) $\gamma\approx0$.
Let us note that $\gamma=0$ in the nonrelativistic limit. Finally for
$\mathit{\Sigma}_{\pi\text{N}}=73$ MeV the mass of $\Xi_{3/2}$ corresponds to
the estimate of NA49 \cite{Xi}. This is also the value preferred by the recent
analysis of $\pi\text{N}$ scattering \cite{Sigma}.

For eikosiheptaplet the masses (in MeV) are listed in the Table \ref{27mass}.
\begin{table}[h]
\caption{Masses of eikosiheptaplets for different values of $\mathit{\Sigma
}_{\pi\text{N}}$}%
\label{27mass}
\begin{tabular}
[c]{|c|cc|cc|cc|}\hline
$\mathit{\Sigma}_{\pi\text{N}}$ & \multicolumn{2}{c}{42~MeV} &
\multicolumn{2}{|c|}{55~MeV} & \multicolumn{2}{c|}{73~MeV}\\\hline
\text{Spin} & 3/2 & 1/2 & 3/2 & 1/2 & 3/2 & 1/2\\\hline
$\Theta$ & 1568 & 1999 & 1578 & 1965 & 1593 & 1919\\
$\Delta$ & 1721 & 2213 & 1688 & 2165 & 1642 & 2098\\
N & 1715 & 2158 & 1717 & 2087 & 1721 & 1988\\
$\Gamma$ & 1875 & 2439 & 1798 & 2365 & 1691 & 2264\\
$\Sigma$ & 1866 & 2358 & 1837 & 2261 & 1796 & 2126\\
$\Lambda$ & 1862 & 2318 & 1856 & 2209 & 1850 & 2058\\
$\Xi_{3/2}$ & 2018 & 2558 & 1956 & 2435 & 1872 & 2264\\
$\Xi$ & 2011 & 2521 & 1986 & 2399 & 1951 & 2230\\
$\Omega$ & 2160 & 2677 & 2115 & 2504 & 2052 & 2265\\\hline
\end{tabular}
\end{table}

Table \ref{27mass} deserves a few comments. The first two columns
corresponding to $\mathit{\Sigma}_{\pi\text{N}}=42$ MeV are in agreement with
the numerical values from Ref.\cite{Borisyuk1} where N$^{\ast}(1710)$ was
taken as input. The last two columns corresponding to the antidecuplet masses:
$M_{\Theta^{+}}=1535$ and $M_{\Xi_{3/2}}=1860$ are in agreement with
Refs.\cite{Ma1,Borisyuk2,Ma2}. Finally, let us observe that -- as can be also
seen from Fig.1 -- the spin 1/2 eikosiheptaplet is squeezed for smaller values
of the hypercharge making the heaviest isospin submultiplets almost
degenerate. On the other hand $\Theta_{27}$ in $27_{3/2}$ is only a few tens
of GeV above the $\Theta^{+}$ of antidecuplet.

\vspace{0.3cm} \begin{figure}[h]
\includegraphics[scale=0.9]{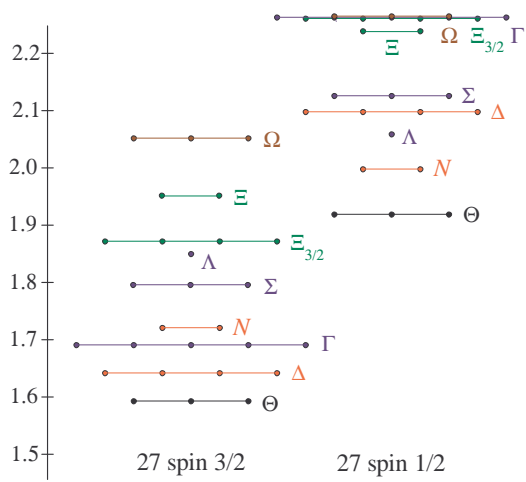}\caption{Spectrum of eikosiheptaplet (in
GeV) of spin 3/2 (left) and spin 1/2 (right) for $\Sigma_{\pi\text{N}}%
=73$~MeV. Note large splittings of equal hypercharge multiplets.}%
\label{fig:eiko}%
\end{figure}

\section{Decay constants for decuplet and antidecuplet}

The matrix elements for decuplet and antidecuplet with $S_{3}=S_{3}^{\prime
}=1/2$ read:%
\begin{align}
\mathcal{A}(B_{10_{3/2}}  &  \rightarrow B_{8}^{\prime}+\varphi)=3\left(
\begin{array}
[c]{cc}%
8 & 8\\
\varphi & B^{\prime}%
\end{array}
\right\vert \left.
\begin{array}
[c]{c}%
10\\
B
\end{array}
\right)  \frac{2}{\sqrt{15}}\times G_{10},\label{dec10}\\
\mathcal{A}(B_{\overline{10}_{1/2}}  &  \rightarrow B_{8}^{\prime}%
+\varphi)=-3\left(
\begin{array}
[c]{cc}%
8 & 8\\
\varphi & B^{\prime}%
\end{array}
\right\vert \left.
\begin{array}
[c]{c}%
\overline{10}\\
B
\end{array}
\right)  \frac{1}{\sqrt{15}}\times G_{\overline{10}}, \label{deca10}%
\end{align}
where%
\begin{equation}
G_{10}=G_{0}+\frac{1}{2}G_{1},\;G_{\overline{10}}=G_{0}-G_{1}-\frac{1}{2}%
G_{2}. \label{G10a10}%
\end{equation}

In order to have an estimate of the width (\ref{gammadef}) the authors of
Ref.\cite{DPP} calculated $G_{\overline{10}}$ in the nonrelativistic limit of
$\chi$QSM \cite{limit} and got $G_{\overline{10}}\equiv0$. It has been shown
\cite{MPwidth} that this cancelation between terms that scale differently with
$N_{c}$ ($G_{0}\sim N_{c}^{3/2},\,G_{1,2}\sim N_{c}^{1/2}$) is actually
consistent with large $N_{c}$ counting, since in fact%
\begin{equation}
G_{\overline{10}}=G_{0}-\frac{N_{c}+1}{4}G_{1}-\frac{1}{2}G_{2}%
\end{equation}
where the explicit $N_{c}$ dependence comes from the SU(3) Clebsch-Gordan
coefficients calculated for large $N_{c}$ (note that for arbitrary $N_{c}$
baryons are built from $N_{c}$ quarks rather than from 3). In the
nonrelativistic limit (NRL) \cite{MPwidth}:%
\begin{equation}
G_{0}=-(N_{c}+2)\,G,~ G_{1}=-4G, ~G_{2}=-2G,~ G\sim N_{c}^{1/2}. \label{NREL}%
\end{equation}
Similar cancelations occur also for the decays of the eikosiheptaplet
\cite{Piesciuk}. From now on we will keep $N_{c}=3$.

Following steps described in the Appendix we obtain the averaged matrix
elements%
\begin{align}
\overline{\mathcal{A}^{2}}(B_{10_{3/2}}  &  \rightarrow B_{8}^{\prime}%
+\varphi)=\frac{6}{5}\left[
\begin{array}
[c]{cc}%
8 & 8\\
\varphi & B^{\prime}%
\end{array}
\right\vert \left.
\begin{array}
[c]{c}%
10\\
B
\end{array}
\right]  ^{2}\times G_{10}^{2},\label{decsof10}\\
\overline{\mathcal{A}^{2}}(B_{\overline{10}_{1/2}}  &  \rightarrow
B_{8}^{\prime}+\varphi)=\frac{3}{5}\left[
\begin{array}
[c]{cc}%
8 & 8\\
\varphi & B^{\prime}%
\end{array}
\right\vert \left.
\begin{array}
[c]{c}%
\overline{10}\\
B
\end{array}
\right]  ^{2}\times G_{\overline{10}}^{2} \label{decsofa10}%
\end{align}
where the squares of the isoscalar factors (the quantities in the square
brackets in Eqs.(\ref{decsof10},\ref{decsofa10})) are listed in Table
\ref{iso10a10}.

\begin{table}[h]
\caption{Isoscalar factors squared for the decays of decuplet and
antidecuplet}%
\label{iso10a10}
\begin{tabular}
[c]{|l|l|c|}\hline
$10\rightarrow8+8 $ & $\overline{10}\rightarrow8+8 $ & $C^{2}$\\\hline
$\Omega\rightarrow\overline{K}+\Xi$ & $\Theta\rightarrow$K$+$N & $1$\\\hline
$\Xi^{\ast}\rightarrow\pi+\Xi$ & $N\rightarrow\pi+$N & $1/4$\\
$\Xi^{\ast}\rightarrow\eta+\Xi$ & $N\rightarrow\eta+$N & $1/4$\\
$\Xi^{\ast}\rightarrow\overline{\text{K}}+\Lambda$ & N$\rightarrow
\text{K}+\Lambda$ & $1/4$\\
$\Xi^{\ast}\rightarrow\overline{\text{K}}+\Sigma$ & N$\rightarrow
\text{K}+\Sigma$ & $1/4$\\\hline
$\Sigma^{\ast}\rightarrow\overline{\text{K}}+$N & $\Sigma\rightarrow
\overline{\text{K}}+\text{N} $ & $1/6$\\
$\Sigma^{\ast}\rightarrow\pi+\Lambda$ & $\Sigma\rightarrow\pi+\Lambda$ &
$1/4$\\
$\Sigma^{\ast}\rightarrow\pi+\Sigma$ & $\Sigma\rightarrow\pi+\Sigma$ & $1/6
$\\
$\Sigma^{\ast}\rightarrow\eta+\Sigma$ & $\Sigma\rightarrow\eta+\Sigma$ &
$1/4$\\
$\Sigma^{\ast}\rightarrow\text{K}+\Xi$ & $\Sigma\rightarrow\text{K}+\Xi$ &
$1/6$\\\hline
$\Delta\rightarrow\pi+$N & $\Xi_{3/2}\rightarrow\pi+\Xi$ & $1/2$\\
$\Delta\rightarrow\text{K}+\Sigma$ & $\Xi_{3/2}\rightarrow\overline{\text{K}%
}+\Sigma$ & $1/2$\\\hline
\end{tabular}
\end{table}

In Fig.1 we plot \emph{scaled} coupling constants $G_{10}$ and $G_{\overline
{10}}$ (\emph{i.e.} without Goldberger-Treiman factors $(M+M^{\prime
})/3f_{\varphi}$ (\ref{GT})) as functions of parameter $\rho$, where $\rho$ is
given by eq.(\ref{inputa12}). As already explained in the Introduction
$G_{10}$ is constant, as the $\rho$ dependence cancels out, while
$G_{\overline{10}}$ steeply decreases reaching zero for $\rho_{0}=-1.897$.
This is a reflection of the nonrelativistic cancelation (\ref{NREL}) observed
for the first time in Ref.\cite{DPP}. It is obvious that by an appropriate
choice of $\rho$ in the vicinity of $\rho_{0}$ we can make $G_{\overline{10}}$
arbitrarily small. By plugging in parameters (\ref{inputa},\ref{inputa12})
into (\ref{GT}) and (\ref{G10a10}) we get that%
\begin{equation}
\mathit{\Gamma}_{\Theta}<1\,\text{MeV\thinspace}\rightarrow\rho_{1}%
=-1.98<\rho<\rho_{2}=-1.814. \label{Gamrho}%
\end{equation}
In Table \ref{a10decays} we list the decay widths for the remaining members of
antidecuplet for $\rho=-1.98$ (or equivalently $-1.814$) for various choices
of the masses from Table \ref{a10mass} parameterized by the pion-nucleon sigma
term $\Sigma_{\pi\text{N}}$:

\begin{table}[ptb]
\caption{Decay widths in MeV for the decays of antidecuplet}%
\label{a10decays}
\begin{tabular}
[c]{|l|ccc|}\hline
$B_{\overline{10}}\rightarrow\varphi+B_{8}^{\prime} $ &
\multicolumn{3}{c|}{$\mathit{\Gamma}_{B\rightarrow\varphi+B^{\prime}}$
[MeV]}\\\hline
$\mathit{\Sigma}_{\pi\text{N}}$~[MeV] & 42 & 55 & 73\\\hline\hline
$\Theta\rightarrow K+$N & {0.95} & {0.95} & {0.95}\\\hline
N$\rightarrow\pi+$N & {4.18} & {3.77} & {3.25}\\
N$\rightarrow\eta+$N & {0.99} & {0.80} & {0.56}\\
N$\rightarrow K+\Lambda$ & {0.24} & {0.14} & {0.04}\\
N$\rightarrow K+\Sigma$ & {0.02} & {$-$} & {$-$}\\\hline
$\Sigma\rightarrow\overline{K}+N $ & {1.95} & {1.53} & {1.04}\\
$\Sigma\rightarrow\pi+\Lambda$ & {4.40} & {3.57} & {2.59}\\
$\Sigma\rightarrow\pi+\Sigma$ & {2.24} & {1.77} & {1.22}\\
$\Sigma\rightarrow\eta+\Sigma$ & {0.54} & {0.25} & {0.01}\\
$\Sigma\rightarrow K+\Xi$ & {0.10} & {0.01} & {$-$}\\\hline
$\Xi_{3/2}\rightarrow\pi+\Xi$ & {8.41} & {6.01} & {3.44}\\
$\Xi_{3/2}\rightarrow\overline{K}+\Sigma$ & {4.52} & {2.89} & {1.20}\\\hline
\end{tabular}
\end{table}

We see from Table \ref{a10decays} that the widths of cryptoexotic nucleon and
$\Sigma$ resonances exceed 1 MeV, the width of $\Xi_{3/2}$ is even larger,
however within the limits set by NA49. It is important to observe that the
estimate from Ref.\cite{Borisyuk2} is almost 4 times bigger; it is difficult
to comment why because the authors of Ref.\cite{Borisyuk2} give no details of
their width calculation. One has to remember that the entries in Table
\ref{a10decays} constitute in fact the \emph{upper} limits, since the widths
scale as $(\rho-\rho_{0})^{2}$ (with $\rho_{0}=-1.897$), and can be
arbitrarily decreased with an appropriate choice of $\rho$. In the situation
when the leading contributions are small, $m_{s}$ corrections become
important, that issue has been studied in Ref.\cite{MPbreaking}.

\section{Decay constants for eikosiheptaplet}

In this Section we shall consider decays of eikosiheptaplet ($27$) that can
have either spin $1/2$ or $3/2$, the latter being lighter. Matrix elements for
the decays of eikosiheptaplet of $S=3/2$ (and with $S_{3}=1/2$) read:%
\begin{align}
\mathcal{A}(B_{27_{3/2}}  &  \rightarrow B_{8}^{\prime}+\varphi)=3\left(
\begin{array}
[c]{cc}%
8 & 8\\
\varphi & B^{\prime}%
\end{array}
\right\vert \left.
\begin{array}
[c]{c}%
27\\
B
\end{array}
\right)  \frac{2\sqrt{2}}{9}\times G_{27},\nonumber\\
\mathcal{A}(B_{27_{3/2}}  &  \rightarrow B_{10}^{\prime}+\varphi)=-3\left(
\begin{array}
[c]{cc}%
8 & 10\\
\varphi & B^{\prime}%
\end{array}
\right\vert \left.
\begin{array}
[c]{c}%
27\\
B
\end{array}
\right)  \frac{\sqrt{10}}{36}\times F_{27},\nonumber\\
\mathcal{A}(B_{27_{3/2}}  &  \rightarrow B_{\overline{10}}^{\prime}%
+\varphi)=3\left(
\begin{array}
[c]{cc}%
8 & \overline{10}\\
\varphi & B^{\prime}%
\end{array}
\right\vert \left.
\begin{array}
[c]{c}%
27\\
B
\end{array}
\right)  \frac{\sqrt{30}}{9}\times E_{27},
\end{align}
where%
\begin{align}
G_{27}  &  =G_{0}-\frac{1}{2}G_{1},\nonumber\\
F_{27}  &  =G_{0}-\frac{1}{2}G_{1}-\frac{3}{2}G_{2},\nonumber\\
E_{27}  &  =G_{0}+G_{1}. \label{coup273}%
\end{align}

For $S=1/2$ and $S_{3}=1/2$ we have:%
\begin{align}
\mathcal{A}(B_{27_{1/2}}  &  \rightarrow B_{8}^{\prime}+\varphi)=-3\left(
\begin{array}
[c]{cc}%
8 & 8\\
\varphi & B^{\prime}%
\end{array}
\right\vert \left.
\begin{array}
[c]{c}%
27\\
B
\end{array}
\right)  \frac{\sqrt{10}}{45}\times H_{27},\nonumber\\
\mathcal{A}(B_{27_{1/2}}  &  \rightarrow B_{10}^{\prime}+\varphi)=-3\left(
\begin{array}
[c]{cc}%
8 & 10\\
\varphi & B^{\prime}%
\end{array}
\right\vert \left.
\begin{array}
[c]{c}%
27\\
B
\end{array}
\right)  \frac{\sqrt{2}}{9}\times G_{27}^{\prime},\nonumber\\
\mathcal{A}(B_{27_{1/2}}  &  \rightarrow B_{\overline{10}}^{\prime}%
+\varphi)=3\left(
\begin{array}
[c]{cc}%
8 & \overline{10}\\
\varphi & B^{\prime}%
\end{array}
\right\vert \left.
\begin{array}
[c]{c}%
27\\
B
\end{array}
\right)  \frac{7\sqrt{2}}{36}\times H_{27}^{\prime},
\end{align}
where%
\begin{align}
{H_{27}}  &  {=}{G}_{0}{-}2{G_{1}+\frac{3}{2}G_{2},}\nonumber\\
G_{27}^{\prime}  &  ={G}_{0}{-2G_{1},}\nonumber\\
H_{27}^{\prime}  &  =G_{0}+\frac{11}{14}G_{1}+\frac{3}{14}G_{2}.
\label{coup271}%
\end{align}

\begin{figure}[ptb]
\includegraphics[scale=0.6]{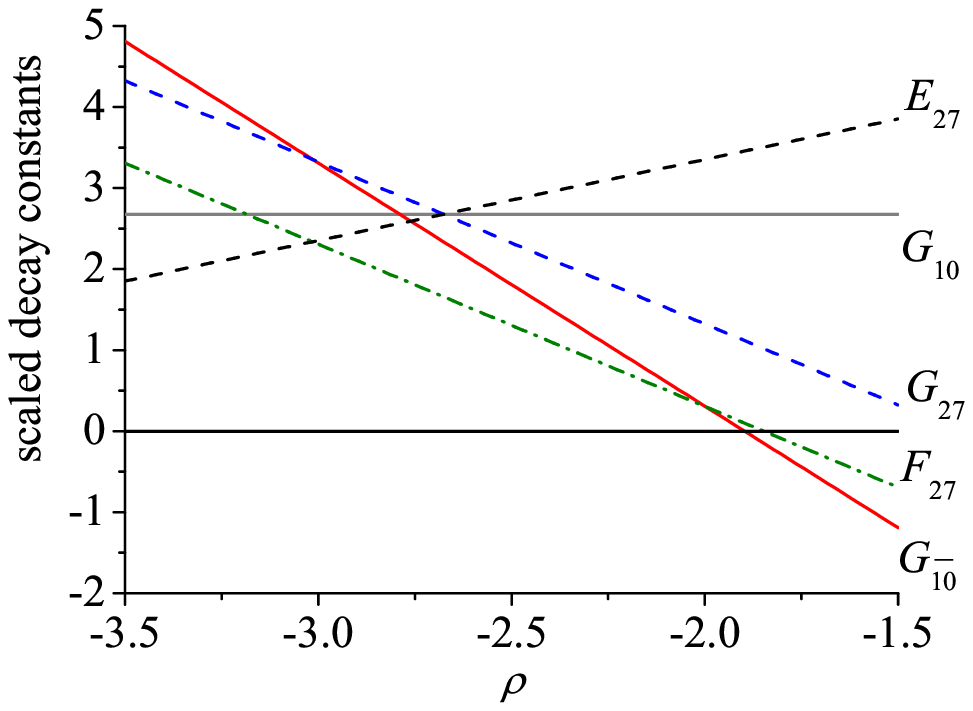}
\includegraphics[scale=0.6]{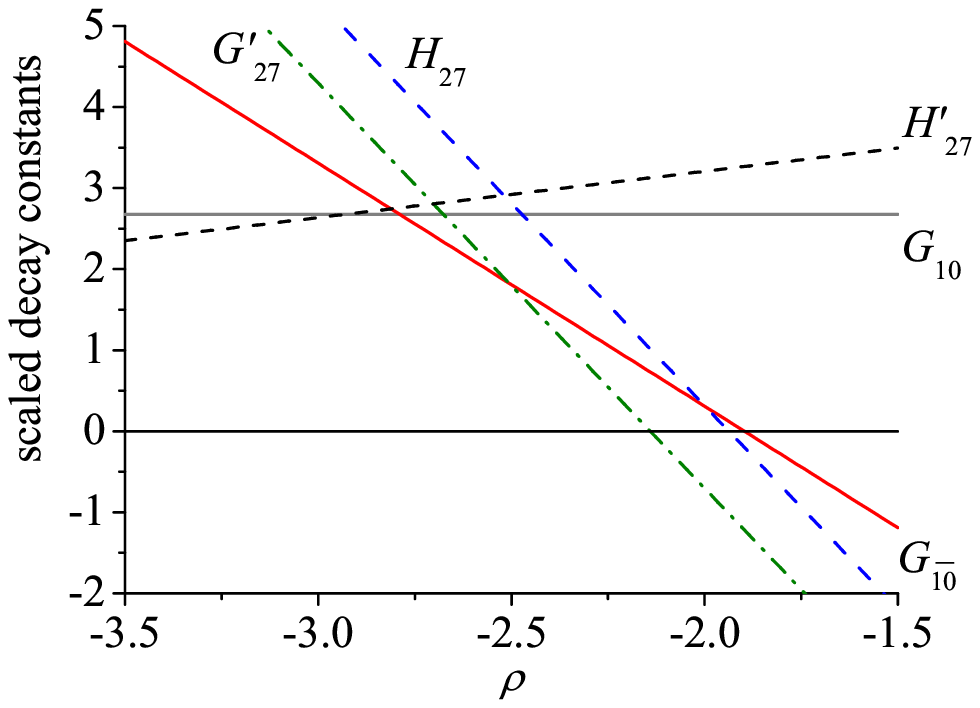}\caption{{{\ Scaled coupling constants
$G_{10}$ and $G_{\overline{10}}$ together with couplings of $27_{3/2}$ (first
panel) and $27_{1/2}$ (second panel) defined in Eqs.(\ref{coup273}%
,\ref{coup271}) as functions of parameter $\rho$, where $\rho$ is given by
eq.(\ref{inputa12}). }}}%
\label{fig:spin3}%
\end{figure}

In Fig.2 we plot \emph{scaled} coupling constants (\emph{i.e.} without
Goldberger-Treiman factors $(M+M^{\prime})/3f_{\varphi}$ (\ref{GT})) for
decays of $27_{3/2}$ and $27_{1/2}$ together with $G_{10}$ and $G_{\overline
{10}}$ (solid lines) as functions of parameter $\rho$, where $\rho$ is given
by eq.(\ref{inputa12}). Together with aforementioned suppression of
$G_{\overline{10}}$ we see strong suppression of $F_{27}$ (corresponding to
$27_{3/2}\rightarrow10_{3/2}+\varphi$) and $H_{27}$ (corresponding to
$27_{1/2}\rightarrow8_{1/2}+\varphi$) for the same range of $\rho$.
Interestingly, both $F_{27}$ and $H_{27}$ vanish \cite{Piesciuk} in the
nonrelativistic limit (\ref{NREL}) exactly as $G_{\overline{10}}$. In our
parametrization they cross zero for the parameter $\rho$ in the range
(\ref{Gamrho}). Somewhat smaller suppression is seen for spin changing
transitions $G_{27}$ (corresponding to $27_{3/2}\rightarrow8_{1/2}+\varphi$)
and $G_{27}^{\prime}$ (corresponding to $27_{1/2}\rightarrow10_{3/2}+\varphi
$). Interestingly, in the nonrelativistic limit there is a partial cancelation
in these couplings, namely the leading $N_{c}$ coefficients cancel out
\cite{Piesciuk}. Finally, the remaining couplings $E_{27}$ (corresponding to
$27_{3/2}\rightarrow\overline{10}_{1/2}+\varphi$) and $H_{27}^{\prime}$
(corresponding to $27_{1/2}\rightarrow\overline{10}_{1/2}+\varphi$) are not
suppressed (they are neither suppressed in the nonrelativistic limit).
However, decays to antidecuplet have much smaller phase space, and they are
totally switched off for $27_{3/2}$. It is remarkable that our simple
phenomenological parametrization (\ref{inputa},\ref{inputa12}) respects -- for
the $\rho$ values of interest (\ref{Gamrho}) -- the large $N_{c}$ suppression
in the nonrelativistic limit.

Averaging over spin and isospin, as described in the Appendix gives:
\begin{align}
\overline{\mathcal{A}^{2}}(B_{27_{3/2}}  &  \rightarrow B_{8}^{\prime}%
+\varphi)=\frac{4}{9}\left[
\begin{array}
[c]{cc}%
8 & 8\\
\varphi & B^{\prime}%
\end{array}
\right\vert \left.
\begin{array}
[c]{c}%
27\\
B
\end{array}
\right]  ^{2}\times G_{27}^{2},\nonumber\\
\overline{\mathcal{A}^{2}}(B_{27_{3/2}}  &  \rightarrow B_{10}^{\prime
}+\varphi)=\frac{25}{72}\left[
\begin{array}
[c]{cc}%
8 & 10\\
\varphi & B^{\prime}%
\end{array}
\right\vert \left.
\begin{array}
[c]{c}%
27\\
B
\end{array}
\right]  ^{2}\times F_{27}^{2},\nonumber\\
\overline{\mathcal{A}^{2}}(B_{27_{3/2}}  &  \rightarrow B_{\overline{10}%
}^{\prime}+\varphi)=\frac{5}{3}\left[
\begin{array}
[c]{cc}%
8 & \overline{10}\\
\varphi & B^{\prime}%
\end{array}
\right\vert \left.
\begin{array}
[c]{c}%
27\\
B
\end{array}
\right]  ^{2}\times E_{27}^{2}, \label{decsof3}%
\end{align}
where the quantities in the square brackets denote $SU(3)$ isoscalar factors.
For $27_{1/2}$ we get%
\begin{align}
\overline{\mathcal{A}^{2}}(B_{27_{1/2}}  &  \rightarrow B_{8}^{\prime}%
+\varphi)=\frac{2}{45}\left[
\begin{array}
[c]{cc}%
8 & 8\\
\varphi & B^{\prime}%
\end{array}
\right\vert \left.
\begin{array}
[c]{c}%
27\\
B
\end{array}
\right]  ^{2}\times H_{27}^{2},\nonumber\\
\overline{\mathcal{A}^{2}}(B_{27_{1/2}}  &  \rightarrow B_{10}^{\prime
}+\varphi)=\frac{2}{9}\left[
\begin{array}
[c]{cc}%
8 & 10\\
\varphi & B^{\prime}%
\end{array}
\right\vert \left.
\begin{array}
[c]{c}%
27\\
B
\end{array}
\right]  ^{2}\times G_{27}^{\prime\,2},\nonumber\\
\overline{\mathcal{A}^{2}}(B_{27_{1/2}}  &  \rightarrow B_{\overline{10}%
}^{\prime}+\varphi)=\frac{49}{72}\left[
\begin{array}
[c]{cc}%
8 & \overline{10}\\
\varphi & B^{\prime}%
\end{array}
\right\vert \left.
\begin{array}
[c]{c}%
27\\
B
\end{array}
\right]  ^{2}\times H_{27}^{\prime\,2}. \label{decsof1}%
\end{align}
The squares of the relevant SU(3) isoscalar factors are listed in Table
\ref{27iso}.

\begin{table}[ptb]
\caption{Isoscalar factors squared for the decays of eikosiheptaplet}%
\label{27iso}
\begin{tabular}
[c]{|lc|lc|lc|}\hline
$27\rightarrow8+8$ & $C^{2} $ & $27\rightarrow8+10$ & $C^{2} $ &
$27\rightarrow8+\overline{10}$ & $C^{2}$\\\hline
&  &  &  & $\Theta\rightarrow\pi+\Theta$ & 3/4\\
$\Theta\rightarrow$K$+$N & 1 & $\Theta\rightarrow$K$+\Delta$ & 1 &
$\Theta\rightarrow$K$+$N & 1/4\\\hline
N$\rightarrow\eta+$N & 9/20 &  &  & N$\rightarrow\eta+$N & 9/80\\
N$\rightarrow\pi+$N & 1/20 & N$\rightarrow\pi+\Delta$ & 1/5 & N$\rightarrow
\pi+$N & 49/80\\
N$\rightarrow$K$+\Sigma$ & 1/20 & N$\rightarrow$K$+\Sigma$ & 4/5 &
N$\rightarrow$K$+\Sigma$ & 1/20\\
N$\rightarrow$K$+\Lambda$ & 9/20 &  &  & N$\rightarrow\overline{\text{K}%
}+\Theta$ & 9/40\\\hline
&  & $\Delta\rightarrow\eta+\Delta$ & 9/16 &  & \\
$\Delta\rightarrow\pi+$N & 1/2 & $\Delta\rightarrow\pi+\Delta$ & 5/16 &
$\Delta\rightarrow\pi+$N & 1/2\\
$\Delta\rightarrow$K$+\Sigma$ & 1/2 & $\Delta\rightarrow$K$+\Sigma$ & 1/8 &
$\Delta\rightarrow$K$+\Sigma$ & 1/2\\\hline
\end{tabular}
\end{table}

Now we are ready to calculate the decay widths for $27_{3/2}$. In fact only
decays to the octet baryons have nonvanishing widths, we list them in the
Table \ref{Gam1} ("$\sim0$" denotes the decay width below 1 MeV, whereas "$-$"
means that the decay is kinematically forbidden).

\begin{table}[ptb]
\caption{Decay widths in MeV for the decays of $27_{3/2}$.}%
\label{Gam1}
\begin{tabular}
[c]{|l|rrr|rrr|}\hline
$27_{3/2}\rightarrow8+8$ & \multicolumn{3}{c|}{$\mathit{\Gamma}${~[MeV]}} &
\multicolumn{3}{c|}{$\mathit{\Gamma}${~[MeV]}}\\
$\rho$ & \multicolumn{3}{c|}{$\rho_{1}$} & \multicolumn{3}{c|}{$\rho_{2}$}\\
$\mathit{\Sigma}_{\pi\text{N}}$ & 42 & 55 & 73 & 42 & 55 & 73\\\hline\hline
$\Theta_{27}\rightarrow K+$N & 29 & 33 & 39 & 16 & 18 & 21\\\hline
N$_{27}\rightarrow\eta+$N & 37 & 37 & 38 & 20 & 20 & 21\\
N$_{27}\rightarrow\pi+$N & 17 & 17 & 17 & 9 & 9 & 9\\
N$_{27}\rightarrow$K$+\Sigma$ & $\sim0$ & $\sim0$ & $\sim0$ & $\sim0$ &
$\sim0$ & $\sim0$\\
N$_{27}\rightarrow$K$+\Lambda$ & 10 & 10 & 10 & 5 & 5 & 6\\
$\Delta_{27}\rightarrow\pi+$N & 172 & 152 & 128 & 94 & 83 & 70\\
$\Delta_{27}\rightarrow$K$+\Sigma$ & 2 & $-$ & $-$ & 1 & $- $ & $-$\\\hline
\end{tabular}
\end{table}

Decays of $27_{3/2}$ to decuplet are kinematically forbidden except of the
decays N$_{27} \rightarrow\pi+\Delta$ and $\Delta_{27} \rightarrow\pi+\Delta$
which have widths smaller than 1 MeV. All decays to antidecuplet are
kinematically forbidden. We can therefore conclude that eikosiheptaplet of
spin $3/2$ has widths small enough to justify the rigid rotor quantization.
Not only are the widths numerically smaller than the one of $\Delta$, but also
in the large $N_{c}$ limit with the partial nonrelativistic cancelation taking
place, $\mathit{\Gamma}_{27_{3/2}\rightarrow8+8}\rightarrow0$ \cite{Piesciuk}.

Our results for $\Theta_{27}$ presented in Table \ref{Gam1} are smaller than
the estimate of Ref.\cite{Borisyuk2}. Although the widths of the order of tens
of MeV can be considered small, one has to remember that partial wave analysis
requires $\Delta_{27}$ and $\Theta_{27}$ widths to be of the order of 100 keV
\cite{Azimov}.

We have concentrated here on the lightest states of eikosiheptaplet that have
been looked for in PW analysis \cite{Azimov}. Obviously, we can easily
calculate widths for the plethora of the remaining states of eikosiheptaplet.
We have checked that for other states widths are smaller than the one of
$\Delta_{27}$ quoted above. Assuming $\mathit{\Sigma}_{\pi\text{N}}=73$ MeV we
get the following upper bounds for the partial widths of the next isospin
multiplets%
\begin{align}
\mathit{\Gamma}_{\Lambda\rightarrow\eta+\Lambda}  &  \sim42\;\text{MeV}%
,\nonumber\\
\mathit{\Gamma}_{\Sigma\rightarrow\pi+\Lambda}  &  \sim75\;\text{MeV}%
,\nonumber\\
\mathit{\Gamma}_{\Xi_{1/2}\rightarrow\overline{\text{K}}+\Lambda}  &
\sim68\;\text{MeV},\nonumber\\
\mathit{\Gamma}_{\Xi_{3/2}\rightarrow\pi+\Xi}  &  \sim74\;\text{MeV}.
\end{align}

\begin{table}[ptb]
\caption{Decay widths in MeV for the decays of $27_{1/2}$ to octet.}%
\label{271to8}
\begin{tabular}
[c]{|l|rrr|rrr|}\hline
$27_{1/2}\rightarrow8+8$ & \multicolumn{3}{c|}{$\mathit{\Gamma}${~[MeV]}} &
\multicolumn{3}{c|}{$\mathit{\Gamma}${~[MeV]}}\\
$\rho$ & \multicolumn{3}{c|}{$\rho_{1}$} & \multicolumn{3}{c|}{$\rho_{2}$}\\
$\mathit{\Sigma}_{\pi\text{N}}$ & 42 & 55 & 73 & 42 & 55 & 73\\\hline\hline
$\Theta_{27}\rightarrow$K$+$N & 0.97 & 0.87 & 0.73 & 8.14 & 7.26 &
6.12\\\hline
N$_{27}\rightarrow\eta+$N & 0.69 & 0.56 & 0.40 & 5.83 & 4.70 & 3.36\\
N$_{27}\rightarrow\pi+$N & 0.16 & 0.13 & 0.11 & 1.34 & 1.14 & 0.89\\
N$_{27}\rightarrow$K$+\Sigma$ & 0.04 & 0.03 & 0.02 & 0.32 & 0.24 & 0.14\\
N$_{27}\rightarrow$K$+\Lambda$ & 0.44 & 0.34 & 0.22 & 3.68 & 2.85 &
1.87\\\hline
$\Delta_{27}\rightarrow\pi+$N & 1.80 & 1.62 & 1.34 & 15.12 & 13.60 & 11.66\\
$\Delta_{27}\rightarrow$K$+\Sigma$ & 0.46 & 0.39 & 0.30 & 3.89 & 3.26 &
2.48\\\hline
\end{tabular}
\end{table}

For $27_{1/2}$ we expect larger widths because the available phase-space is
much larger. Interestingly, this is not the case for the decays to octet. The
reason is that $H_{27}$ responsible for these decays is strongly suppressed in
the relevant range of $\rho$. Indeed, $H_{27}$ crosses zero at $\rho=-1.937$
\emph{i.e.} within the range (\ref{Gamrho}). Moreover, the overall group
theoretical factor in eq.(\ref{decsof1}) is suppressed by factor of 13 with
respect to the decays of antidecuplet (\ref{decsofa10}). These two
suppressions overcome the increase of the phase-space volume and the decay
widths are comparable to those of $\overline{10}_{1/2}$. Similar effect takes
place for the decays to decuplet (although the decay constant $G_{27}^{\prime
}$ does not cross zero in the relevant range of $\rho$) and the decays are
comparable to those of $27_{3/2}\rightarrow8+8$. \begin{table}[ptb]
\caption{Decay widths in MeV for the decays of $27_{1/2}$ to decuplet.}%
\label{271to10}
\begin{tabular}
[c]{|l|rrr|rrr|}\hline
$27_{1/2}\rightarrow8+10$ & \multicolumn{3}{c|}{$\mathit{\Gamma}${~[MeV]}} &
\multicolumn{3}{c|}{$\mathit{\Gamma}${~[MeV]}}\\
$\rho$ & \multicolumn{3}{c|}{$\rho_{1}$} & \multicolumn{3}{c|}{$\rho_{2}$}\\
$\mathit{\Sigma}_{\pi\text{N}}$ & 42 & 55 & 73 & 42 & 55 & 73\\\hline\hline
$\Theta_{27}\rightarrow$K$+\Delta$ & 21 & 17 & 12 & 86 & 69 & 48\\\hline
N$_{27}\rightarrow\pi+\Delta$ & 24 & 19 & 14 & 97 & 78 & 56\\
N$_{27}\rightarrow$K$+\Sigma^{\ast}$ & 18 & 11 & 4 & 75 & 45 & 16\\\hline
$\Delta_{27}\rightarrow\eta+\Delta$ & 30 & 24 & 18 & 126 & 103 & 74\\
$\Delta_{27}\rightarrow\pi+\Delta$ & 43 & 37 & 31 & 178 & 155 & 127\\
$\Delta_{27}\rightarrow$K$+\Sigma^{\ast}$ & 4 & 3 & 2 & 16 & 12 & 8\\\hline
\end{tabular}
\end{table}

Unfortunately there is no suppression for the decays of $27_{1/2}$ to
antidecuplet. Indeed, the relevant coupling $H_{27}^{\prime}$ is as large as
$G_{10}$ (responsible for $\Delta$ decay) -- see Fig. \ref{fig:spin3} -- and
the phase space is also not suppressed: for $\Theta_{27}\rightarrow\pi
+\Theta_{\overline{10}}$ the pion momentum is of the order of $300\div400$ MeV
depending on $\mathit{\Sigma}_{\pi{N}}$. Hence the resulting widths are large.
\begin{table}[ptb]
\caption{Decay widths in MeV for the decays of $27_{1/2}$ to antidecuplet.}%
\label{271toa10}
\begin{tabular}
[c]{|l|rrr|rrr|}\hline
$27_{1/2}\rightarrow8+10$ & \multicolumn{3}{c|}{$\mathit{\Gamma}${~[MeV]}} &
\multicolumn{3}{c|}{$\mathit{\Gamma}${~[MeV]}}\\
$\rho$ & \multicolumn{3}{c|}{$\rho_{1}$} & \multicolumn{3}{c|}{$\rho_{2}$}\\
$\mathit{\Sigma}_{\pi\text{N}}$ & 42 & 55 & 73 & 42 & 55 & 73\\\hline\hline
$\Theta_{27}\rightarrow\pi+\Theta_{\overline{10}} $ & 658 & 523 & 365 & 697 &
554 & 387\\
$\Theta_{27}\rightarrow$K$+$N$_{\overline{10}} $ & $-$ & $-$ & $-$ & $-$ & $-$
& $-$\\\hline
N$_{27}\rightarrow\eta+$N$_{\overline{10}} $ & $-$ & $-$ & $-$ & $-$ & $-$ &
$-$\\
N$_{27}\rightarrow\pi+$N$_{\overline{10}}$ & 500 & 364 & 215 & 530 & 385 &
228\\
N$_{27}\rightarrow$K$+\Sigma_{\overline{10}} $ & $-$ & $-$ & $-$ & $-$ & $-$ &
$-$\\
N$_{27}\rightarrow\overline{\text{K}}+\Theta_{\overline{10}} $ & 72 & 20 & $-
$ & 76 & 21 & $-$\\\hline
$\Delta_{27}\rightarrow\pi+$N$_{\overline{10}} $ & 579 & 510 & 424 & 614 &
541 & 449\\
$\Delta_{27}\rightarrow$K$+\Sigma_{\overline{10}} $ & $-$ & $-$ & $-$ & $-$ &
$-$ & $-$\\\hline
\end{tabular}
\end{table}

Therefore one would be tempted to conclude that $27_{1/2}$ cannot be
considered as a semi stable multiplet and its description in terms of the
rigid rotor fails, at least in the situations where the transitions
$27_{1/2}\leftrightarrow$ $\overline{10}$ are of importance. This statement is
however not supported by the $N_{c}$ counting \cite{Piesciuk}. We shall come
back to this issue in the next Section.

\section{Summary}

In the present paper we have studied masses and decay widths of exotic baryon
eikosiheptaplets (\emph{i.e.} 27-plets) of spin 3/2 and 1/2 that follow from
the chiral quark-soliton model in the rigid rotator quantization approach. We
have also reexamined widely studied by now antidecuplet that we use as an
input that constrains model parameters. Rigid rotator quantization predicts a
tower of stable exotic representations of different spins and positive parity,
antidecuplet, eikosiheptaplet, $\overline{35}$ being most prominent examples.
Question arises, where does the rigid rotator approach break? Leaving aside
fundamental problems based on claims in the literature that the rigid rotator
approach to exotica is not compatible with large $N_{c}$ expansion for QCD
\cite{Pobylitsa:2003ju}, we have taken more modest phenomenological approach.
If the widths of the baryonic states calculated within the model exceed
certain critical value, that can be taken to be above the $\Delta$ resonance
width (one has to remember that $\Delta$ can be considered as a well behaved
stable state in the large $N_{c}$ limit), then the model becomes inconsistent.
There are two sources that contribute to the increase of the width with the
increase of the dimensionality of the SU(3) flavor representation. One is
obvious: for higher representations the pertinent states are heavier and the
phase space is larger. The second source is the coupling. For antidecuplet
there is only one coupling corresponding to the transition $\overline
{10}\rightarrow8$, that is excessively small due to the cancelation found in
Ref.\cite{DPP} and discussed in some detail in Sect. 3. For higher
representations there are more couplings corresponding to different
transitions and some of them are not suppressed. For eikosiheptaplet couplings 
to antidecuplet are not suppressed. Obviously if the phase space is large and the
coupling is not suppressed then the widths are large. In other cases one has
to perform explicit calculations to see what is the interplay between the
rising phase space and small coupling.

We have addressed this question by applying the so called
\emph{model-independent approach} \cite{ANW} in which the general group
theoretical structure is taken from the model, while the parameters are fitted
to appropriate data. We have used as an input nonexotic masses and the mass of
$\Theta^{+}$, semi-leptonic decay constants and the assumption that
$\mathit{\Gamma}_{\Theta^{+}}<1$ MeV. The residual freedom was parameterized
by the value of the pion-nucleon sigma term. We have confined our analysis to
eikosiheptaplet (\emph{i.e. }27-plet) that is the only exotic representation
(apart from antidecuplet) appearing in the direct product of two octets. For
this reason eikosiheptaplet might be produced in meson-nucleus scattering and
could subsequently decay to meson-nucleon or meson-hyperon final states.

Our findings can be shortly summarized as follows. Based on group theory
alone, eikosiheptaplet can decay into octet, decuplet and exotic antidecuplet.
However, for $27_{3/2}$ regular octet is kinematically the only allowed
channel (with two exceptions discussed in Sect. 4). Furthermore, transition
$27\rightarrow8$ is governed by a small decay coupling, $G_{27}$. Therefore
eikosiheptaplet of spin 3/2 has widths of the order of a few tens MeV with one
exception, namely $\Delta_{27}$ for which $\mathit{\Gamma}\sim70\div170$ MeV.
For spin 1/2 the situation is different. Decays to octet and decuplet have
small transition couplings and the resulting widths are  small
 (see Tables \ref{271to8} and \ref{271to10}). For the decays to
antidecuplet the coupling is large. Therefore whenever the decay is possible
the widths are of the order of 500 MeV. This might be interpreted as the
signal that the model breaks down and that the assumption that $27_{1/2}$ is
stable cannot be justified phenomenologically.

The situation is, however, more complicated. Since the mass difference
\begin{equation}
\Delta_{27_{1/2}-\overline{10}}=\frac{1}{I_{2}}\sim\mathcal{O}(1/N_{c})
\end{equation}
as calculated from eq.(\ref{Hamiltonian}) is suppressed in the large $N_{c}$
limit, so is the meson momentum (\ref{mom}). Therefore the widths that depend
on the third power of momentum may be suppressed in the large $N_{c}$ limit
despite the fact that they are numerically large. That this is indeed the case
was shown in Ref.\cite{Piesciuk} where \emph{e.g.}%
\begin{equation}
\mathit{\Gamma}_{\Theta_{27}\rightarrow\pi+\Theta_{\overline{10}}}%
\sim\mathcal{O}(1/N_{c}^{2}).
\end{equation}
On the contrary, for the transitions of $27_{1/2}$ to octet which are
numerically suppressed (remember that the pertinent coupling $H_{27}$
(\ref{coup271}) vanishes in the NR limit (\ref{NREL})) the $N_{c}$ scaling is
different, \emph{e.g.} \cite{Piesciuk}:%
\begin{equation}
\mathit{\Gamma}_{\Theta_{27}\rightarrow\text{K}+\text{N}}\sim\mathcal{O}(1).
\end{equation}

Obviously numerical results presented in Sect. IV depend on the choice of
input parameters. We have studied this sensitivity by employing another set of
parameters (\ref{fit2}) that corresponds to more realistic $g_{A}^{(0)}$.
The decay widths of antidecuplet do not change, since we require
that $\mathit{\Gamma}_{\Theta}<1$ (which is equivalent to slightly different
range of the parameter $\rho$: , $\rho_{1}=-1.933<\rho<\rho_{2}=-1.767$) and
this condition fixes all remaining decay widths. For eikosiheptaplet some
differences appear. For the transitions of $27_{3/2}$ to octet the decay
widths for fit 2 (\ref{fit2}) are smaller by a few MeV. More drastic changes
appear for $27_{1/2}$. The reason is that the small change in the coupling is
magnified by a large phase space factor. Indeed, for the decays to octet,
presented in Table VII, the decay widths for fit 2 are larger by a factor
$10\div6$ (first number refers to $\rho=\rho_{1}$ whereas the second one to
$\rho=\rho_{2}$ for fit 2). Although this enhancement seems large, the
absolute values are still small on a typical hadronic scale. Less drastic
enhancement occurs for the decays to decuplet presented in Table VIII, the
decay widths for fit 2 are larger by a factor $2\div1.4$. Finally, large decay
widths to antidecuplet remain almost the same as for the fit 1 (\ref{inputa12}%
). We see therefore that despite some numerical uncertainties due to the
choice of input parameters the general pattern persists and our conclusions
still hold.

Summarizing: there is no simple way to judge the quality of the rigid rotator
approach to the eikosiheptaplet. On the basis of phenomenology alone one would
conclude that $27_{1/2}$ is unstable because of the large numerical values of
the decay widths to antidecuplet. On the other hand precisely these decays are
damped in the large $N_{c}$ limit, similarly to the decays of $\Delta$
resonance. Other decays, like the decays to octet scale as $\mathcal{O}%
(N_{c}^{0})$ but their numerical values are small due to the coupling
suppression and additional group theoretical factors.
For eikosiheptaplet of spin $3/2$ all kinematically allowed decays
have widths small enough to justify the rigid rotor quantization.
Not only are the widths numerically smaller than the one of $\Delta$, but also
in the large $N_{c}$ limit partial nonrelativistic cancelation takes place and
the pertinent couplings are suppressed.

\begin{acknowledgments}
The authors are grateful to Hyun-Chul Kim and Ghil-Seok Yang for useful
discussions. The paper was partially supported by the Polish-German
cooperation agreement between Polish Academy of Science and DFG. K.G. was also
supported by the COSY-J{\"u}lich project.
\end{acknowledgments}

\appendix

\section{Summing over spins and isospins}

We shall use the identity%
\begin{equation}
\frac{1}{2S+1}%
{\displaystyle\sum\limits_{S_{3}^{\prime},S_{3}}}
\left(
\begin{array}
[c]{cc}%
1 & S^{\prime}\\
m^{\prime} & S^{\prime}_{3}%
\end{array}
\right\vert \left.
\begin{array}
[c]{c}%
S\\
S_{3}%
\end{array}
\right)  \left(
\begin{array}
[c]{cc}%
1 & S^{\prime}\\
m & S^{\prime}_{3}%
\end{array}
\right\vert \left.
\begin{array}
[c]{c}%
S\\
S_{3}%
\end{array}
\right)  =\frac{1}{3}\delta_{mm^{\prime}}%
\end{equation}
to average over the initial spin (and in the same time to sum over the final
spin). For spin $3/2$ the amplitude for $10_{3/2}\rightarrow8_{1/2}+\varphi$
and $27_{3/2}\rightarrow8_{1/2},\overline{10}_{1/2}+\varphi$ is proportional
to%
\begin{equation}
\left(
\begin{array}
[c]{cc}%
1 & 1/2\\
0 & -1/2
\end{array}
\right\vert \left.
\begin{array}
[c]{c}%
3/2\\
-1/2
\end{array}
\right)  =\sqrt{\frac{2}{3}}%
\end{equation}
hence%
\begin{equation}
\frac{1}{2S+1}%
{\displaystyle\sum\limits_{S_{3}}}
\left\vert \mathcal{A}(3/2\rightarrow1/2)\right\vert ^{2}=\frac{1}%
{2}\left\vert \mathcal{A}(3/2\rightarrow1/2)\right\vert ^{2}.
\end{equation}
For $27_{3/2}\rightarrow10_{3/2}+\varphi$ the amplitude is proportional to%
\begin{equation}
\left(
\begin{array}
[c]{cc}%
1 & 3/2\\
0 & -1/2
\end{array}
\right\vert \left.
\begin{array}
[c]{c}%
3/2\\
-1/2
\end{array}
\right)  =\sqrt{\frac{1}{15}}%
\end{equation}
and%
\begin{equation}
\frac{1}{2S+1}%
{\displaystyle\sum\limits_{S_{3}}}
\left\vert \mathcal{A}(3/2\rightarrow3/2)\right\vert ^{2}=5\left\vert
\mathcal{A}(3/2\rightarrow3/2\right\vert ^{2}.
\end{equation}
Finally, for $\overline{10}_{1/2}\rightarrow8_{1/2}+\varphi$ and
$27_{1/2}\rightarrow8_{1/2},\overline{10}_{1/2}+\varphi$ the amplitude is
proportional to%
\begin{equation}
\left(
\begin{array}
[c]{cc}%
1 & 1/2\\
0 & -1/2
\end{array}
\right\vert \left.
\begin{array}
[c]{c}%
1/2\\
-1/2
\end{array}
\right)  =\sqrt{\frac{1}{3}}%
\end{equation}
and for $27_{1/2}\rightarrow10_{3/2}+\varphi$ to%
\begin{equation}
\left(
\begin{array}
[c]{cc}%
1 & 3/2\\
0 & -1/2
\end{array}
\right\vert \left.
\begin{array}
[c]{c}%
1/2\\
-1/2
\end{array}
\right)  =-\sqrt{\frac{1}{3}}.
\end{equation}
Hence%
\begin{equation}
\frac{1}{2S+1}%
{\displaystyle\sum\limits_{S_{3}}}
\left\vert \mathcal{A}(1/2\rightarrow1/2,3/2)\right\vert ^{2}=\left\vert
\mathcal{A}(1/2\rightarrow1/2,3/2)\right\vert ^{2}.
\end{equation}

Similarly we shall average over initial isospin and sum over the final isospin
using the formula%
\begin{equation}
\frac{1}{2I+1}%
{\displaystyle\sum\limits_{I_{\varphi\,3},I_{3}^{\prime},I_{3}}}
\left(
\begin{array}
[c]{cc}%
I_{\varphi} & I^{\prime}\\
I_{\varphi\,3} & I_{3}^{\prime}%
\end{array}
\right\vert \left.
\begin{array}
[c]{c}%
I\\
I_{3}%
\end{array}
\right)  ^{2}=1.
\end{equation}

\end{document}